\documentclass[a4paper,10pt]{article}

\title{Flavor Mixing, Quark Masses, Neutrino Masses\\and\\Neutrino Oscillations}
\author{H. Fritzsch\\University of Munich, Physics Department\\Munich,
Germany}

\begin{document}

\maketitle

\begin{abstract}
We discuss first the flavor mixing of the quarks, using the texture zero mass matrices. Then we study a similar model for the mass matrices of the leptons. We are able to relate
the mass eigenvalues of the charged leptons and of the neutrinos to the
mixing
angles and can predict the masses of the neutrinos. We find a normal
hierarchy
- the masses are 0.004 eV, 0.01 eV and 0.05 eV. The atmospheric mixing
angle
is given by the mass ratios of the charged leptons and the neutrinos. We
find
 about 40$^{\circ}$, consistent with the experiments. The mixing element, connecting
the first neutrino with the electron, is predicted to be 0.05. This prediction can soon be checked by the Daya Bay experiment.

\end{abstract}
\bigskip
\bigskip

We observe three lepton-quark families in nature. The first family
consists of
the electron, its neutrino and the $u$- and $d$-quarks. The members of the
second
family are the muon, its neutrino and the $c$- and $s$-quarks. The third family
consists of the tauon, its neutrino and the $t$- and $b$-quarks. It is unknown,
whether there is a connection between the number of families and the
number of colors in quantum chromodynamics.

In the Standard Model 28 fundamental constants have to be introduced. Their
values cannot be calculated - they have to be measured in the experiments.
Theses constants are:
\begin{itemize}
\item the constant of gravity $G$,
\item the fine structure constant,
\item the coupling constant of the weak interactions,
\item the coupling constant of the strong interactions or the scale of QCD,
\item the mass of the W-boson,
\item the mass of the ``Higgs''-boson,
\item the masses of the three charged leptons,
\item the masses of the three neutrinos,
\item the masses of the six quarks,
\item the four parameters, describing the flavour mixing of the quarks and
\item the six parameters, describing the flavour mixing of the leptons.
\end{itemize}
For the masses of the quarks we assume the following values:
\begin{center}
$u: \;5.3 \;\;\;\;\,MeV,$\\
$d: \;7.8 \;\;\;\;\,MeV,$\\
$s: \;146 \;\;\;\,MeV,$\\
$c: \;1050 \;\;\, MeV,$\\
$b: \;4 600 \;\;\;\, MeV,$\\
$t: \;174000\; MeV.$
\end{center}

The quark masses are scale dependent. The values given above, except the
one
for the t-mass, are normalized at an energy scale of 1 GeV. The
experimental
errors are not given, but are at least 10 \%, except for the t-mass,
which is
known to about 1 \%.

The quarks of the same charge do mix. If the u-quark interacts with a
W-boson,
a mixture of d, s and b appears. These mixtures are described by the CKM
matrix\cite{Kobayashi}. I prefer a description, which we introduced some time
ago\cite{Fritzsch}:
\begin{equation}
V=\left[\begin{array}{ccc}c_u&s_u&0\\-s_{{u}}&c_{{u}}&0\\0&0&1\end{array}\right]
\left[\begin{array}{ccc}e^{-i\phi}&0&0\\0&c&s\\0&-s&c\end{array}\right]\left
[\begin{array}{ccc}c_{{d}}&-s_{{d}}&0\\s_{{d}}&c_{{d}}&0\\0&0&1\end{array}
\right]
\end{equation}
In the case of three families there are three mixing angles and one phase
parameter. The latter describes the CP-violation. The angles with the
index $u$
or $d$ describe the mixing in the $u-c$ sector or the $d-s$ sector, the angle
with no index describes the mixing between the $(t,c)$-system and the
$(b,s)$-system.

I proposed years ago a simple texture $0$ mass matrix for the quarks,
given by
the following matrix, valid both for the $(u,c,t)$-system and the
$(d,s,b)$-system\cite{Fritzsch79}:

\begin{equation}
M=\left[ \begin{array}{ccc}0&A&0\\A^{{*}}&C&B\\0&B^{{*}}&D\end{array}\right]
\end{equation}
Such mass matrices are obtained, if in the electroweak theory special
symmetries are present, either discrete reflection symmetries or continuous
symmetries\cite{Fritzsch79}. In the Standard Model such symmetries cannot be
introduced, but if the electroweak gauge group contains besides the
lefthanded group also a righthanded group, such symmetries can
easily be obtained. A left-right symmetric gauge group arises, for
example, if one uses the group $SO(10)$ to describe the grand unification.

After diagonalization one finds the following relations for the mixing
angles\cite{Fritzsch99}:
\begin{equation}
\theta_d=\tan^{-1}\sqrt{m_d/m_s}\;\;\;\;\;,\;\;\;\;\;
\theta_u=\tan^{-1}\sqrt{m_u/m_c}
\end{equation}
For the angle $\theta_d$ we find from the ratio of the quark masses
$13 \pm 0.4^{\circ}$, the experimental value is $11.7 \pm 2.6^{\circ}$.
The angle $\theta_u$ is calculated to $5.0 \pm 0.7^{\circ}$, the experimental
value is $5.4 \pm 1.1^{\circ}$. Thus these angles agree very well with the
experimental data on flavor mixing, if the phase angle is assumed to be
close to $90^{\circ}$. The experiments give for the phase angle
$86 \ldots 95^{\circ}$. Note that in this case the flavor mixing matrix is
very simple - instead of the arbitrary phase the matrix element is just
given by $-i$.

Analogously we can describe the flavour mixing in the lepton sector,
which can
be studied in the neutrino oscillations. The experiments, carried out in
Japan
(Kamiokande detector\cite{Cravans}) and in Canada (SNO detector, ref. 6), give
the following results for the mass-squared differences of the three neutrino
mass
eigenstates:
\begin{equation}
\Delta m_{21}2\simeq 8\times10^{-5} eV^2\;\;\;\;\;,\;\;\;\;\;\Delta
m_{32}2 \simeq 2.5\times10^{-3} eV^2
\end{equation}
In neutrino oscillations only mass squared differences can be measured. No
information can be obtained for the absolute magnitude of the neutrino
masses.
There is the possibility that the neutrino masses are nearly degenerate,
e.g.
masses like $m_1 = 0.94  eV$, $m_2 =   0.95 eV$, $m_3 = 1   eV$
are possible. If one neutrino remains massless, one would have the masses
$m_1 = 0eV$,$m_2=0.009eV$, $m_3 = 0.05  eV$. In the latter case a hierarchy
of the masses is present, but this hierarchy is much weaker than the mass
hierarchy for  the charged leptons.

Below we shall calculate the neutrino
masses. Neutrino oscillations arise, since the neutrinos, produced by the weak
interactions, are not mass eigenstates, but mixtures of mass eigenstates.
Like for the quarks one has a $3 \times 3$ mixing matrix, which can be written
as a product of three simple matrices, and a phase matrix, which is present
only, if neutrinos are Majorana particles:
\begin{center}
${V=U\cdot P}$
\end{center}
\bigskip
\begin{equation}
U=\left[
\begin{array}{ccc}c_l&s_l&0\\-s_l&c_l&0\\0&0&1\end{array}\right]\left
[\begin{array}{ccc}e^{-i\phi}&0&0\\0&c&s\\0&-s&c\end{array}\right]\left
[\begin{array}{ccc}c_\nu&-s_\nu&0\\s_\nu&c_\nu&0\\0&0&1\end{array}\right]
\end{equation}
\begin{center}
$P=\left[\begin{array}{ccc}e^{i\rho}&0&0\\0&e^{i\sigma}&0\\0&0&1\end{array}
\right] \nonumber$
\end{center}
Here $s_\nu$ stands for $\sin {\theta_\nu}$ ($\theta_\nu$ : solar mixing
angle), s  stands for $\sin {\theta}$ ($\theta$: atmospheric mixing
angle),
and $ s_l $  stands for $\sin {\theta_l} $ ($\theta_l$ : reactor  mixing
angle). The latter has  not been measured. The Chooz experiment gives an
upper limit of 13$^{\circ}$ for this angle.

The experimental results for the
solar and the atmospheric mixing angles are\cite{Cravans, Aharmim}:
\begin{equation}
30^o \leq \theta_\nu \leq 39^o\;\;\;\;\;,\;\;\;\;\; 37^o\leq \theta\leq 53^o
\end{equation}
We assume for the lepton mass matrices the same texture 0 pattern as for
the
quarks\cite{Xing}. Thus we obtain the same relations between the mixing angles
and the mass eigenvalues:

\begin{equation}
\tan {\theta_l}=\sqrt{m_e/m_\mu}\simeq0.07\;\;\;\;,\;\;\;\tan{\theta_\nu}=
\sqrt{m_1/m_2}
\end{equation}
Since the solar angle has been measured (we take 33$^{\circ}$), we find
for the
mass ratio of the first two neutrinos:
\begin{equation}
m_{1}/m_{2} \approx 0.42
\end{equation}
The oscillation experiments provide us with the mass squared
differences. The
new relation\cite{Aharmim} allows us to determine the neutrino masses.
We find (in eV):
\begin{equation}
m_1\approx 0.0036...0.0059\;\;\;\;,\;\;\;\;m_2\approx 0.0085...0.0140
\;\;\;\;,\;\;\;\; m_3\approx 0.044...0.058
\end{equation}
These neutrino masses are very small. We observe also that the masses show a
rather
weak hierarchy, but the mass spectrum is not inverted. The first
neutrino has
the smallest mass.

The atmospheric mixing angle is consistent with 45$^{\circ}$. The
parameter $C$ in
eq. (2) might be zero, as originally assumed\cite{Fritzsch79}. But the high
mass of
the $t$-quark does not allow this possibility for the quarks. It might
work for
the leptons. In this case the atmospheric mixing angle is related to the
two
angles, which are given by the corresponding mass ratios:
\begin{equation}
\tan
{\theta_1}=\sqrt{m_\mu/m_\tau}\;\;\;\;,\;\;\;\tan{\theta_2}=\sqrt{m_2/m_3}
\end{equation}
\begin{center}
$\theta_1\approx 14^o\;\;\;\;,\;\;\;\;\theta_2\approx 24^o$
\end{center}

The atmospheric mixing angle is given by the absolute value of the sum
of the
two angles, including a relative phase between the two terms. In order
to get
the direct sum, this phase must be 180$^{\circ}$. In this case we have for
the
atmospheric mixing angle:
\begin{equation}
\theta = \theta_1 + \theta_2 \simeq 38^o
\end{equation}
We cannot obtain a maximal mixing (45$^{\circ}$), but our result is
consistent
with the experiment.

We can predict the matrix element  $V_{3e}$ of the mixing matrix V:
\begin{equation}
V_{3e}=\sin {\theta} \sin {\theta_l}\approx0.707\sqrt{m_e/m_\mu}\approx 0.05
\end{equation}
A matrix element of this magnitude could be observed in the upcoming
reactor neutrino experiments.

If the neutrino masses are Majorana masses, one expects a neutrinoless double
beta decay. The present limit on the Majorana mass is about
0.23 eV\cite{Amaboldi}, if mixing angles are disregarded. We can estimate the
effect. Only the third neutrino would play a role, however it couples to the
electron with
a magnitude, given by $V_{3e}$. Thus we find an upper limit for the effect
of the order of $0.05 \times 0.05 = 0.0025$, which is a factor 100 smaller
than the
observed limit. The experiments on double beta decay must be improved
by a factor 100 in order to see an effect. This seems extremely difficult.


\newpage
\noindent
\begin{thebibliography}{0}
\bibitem{Kobayashi} M. Kobayashi and T. Maskawa, {\it Prog. Theor. Phys.}
{\bf 49}, 652 (1973).
\bibitem{Fritzsch} H. Fritzsch and Z. Xing, {\it Phys. Letters} {\bf B413},
396 (1997);\\
H. Fritzsch and Z. Xing, {\it Phys. Rev.} {\bf  D57}, 594 (1998).
\bibitem {Fritzsch79} H. Fritzsch, {\it  Nucl. Physl.} {\bf  B155},
189 (1979).
\bibitem{Fritzsch99} H. Fritzsch and Z. Xing, {\it Nucl. Phys.}  {\bf
B556},
49(1999).
\bibitem{Cravans} J. P. Cravans {\it  et al.}, arXiv: 0803.4312.
\bibitem{Aharmim} B. Aharmim {\it et al.}, {\it Phys. Rev.} {\bf C72},
055502
(2005)
\bibitem{Xing} H. Fritzsch and Z. Xing, {\it Phys. Lett.} {\bf B634},
514 (2006)
\bibitem{Amaboldi} C. Amaboldi {\it et al.} arXiv: 0802.3439
\end{thebibliography}
\end{document}